\documentclass[aps,pra,preprint,tightenlines,showpacs,showkeys,superscriptaddress]{revtex4}
\usepackage{amsmath}
\usepackage{graphicx}
\begin{document}
\title{A Laser System for the Spectroscopy\\
 of Highly-Charged Bismuth Ions}
\author{S. Albrecht}
\author{S. Altenburg}
\author{C. Siegel}
\author{N. Herschbach}
\author{G. Birkl}
\affiliation{Institut f\"{u}r Angewandte Physik, Technische Universit\"{a}t Darmstadt, Schlossgartenstra\ss e 7, 64289 Darmstadt, Germany}
\date{\today}
\begin{abstract}
We present and characterize a laser system for the spectroscopy on highly-charged $^{209}$Bi$^{82+}$ ions
at a wavelength of $243.87$ nm. 
For absolute frequency stabilization, the laser system is locked to a near-infra-red laser stabilized to a rubidium transition line using a transfer cavity based locking scheme. Tuning of the output frequency 
with high precision is achieved via a tunable rf offset lock. A sample-and-hold technique gives an extended tuning range of several THz in the UV.
This scheme is universally applicable to the stabilization of laser systems at wavelengths not directly accessible to atomic or molecular resonances.
We determine the frequency accuracy of the laser system using Doppler-free absorption spectroscopy of Te$_2$ vapour at $488$ nm.
Scaled to the target wavelength of $244$ nm, we achieve a frequency uncertainty of $\sigma_{244nm}$ = $ 6.14 $ MHz (one standard deviation) over six days of operation. 
\end{abstract}
\maketitle

\section{Introduction}
\label{intro}
The spectroscopic investigation of hydrogen-like atomic systems has developed into an important test bed for research on atomic structure and bound-state quantum electrodynamics (QED). 
Already high-precision measurements on atomic hydrogen confirm QED theory and its predictions to a high degree.
Even more accurate tests of the predictive power of QED can be achieved by increasing either the precision of the measurement or by increasing the contribution of the effect to be measured \cite{stolker}.
The use of highly-charged heavy ions shifts the magnitude of QED effects from the perturbative regime to a region where higher order terms become important and perturbation theory is no longer valid \cite{winters}.
In addition, the energy scale of the $ 1s $ ground state hyperfine structure is shifted to the laser accessible regime above a nuclear charge number of $Z \approx 60 $ and the upper state lifetime enters a regime where acceptable fluorescence rates are expected \cite{vogel}.\\
Based on these considerations, the spectroscopic investigation of the M1 transition of the ground state hyperfine structure of hydrogen-like bismuth ($^{209}$Bi$^{82+}$) at $243.87$ nm has become one of the prime targets for further research along these lines.
Theoretical investigations of the hyperfine transition energy, the lifetime and their respective QED contributions have been performed \cite{borneis,shabaev,finkenbeiner}.
Measurements of the transition energy have been realized on bunches of $^{209}$Bi$^{82+}$-ions travelling at a velocity of $0.6$ c in the experimental storage ring ESR at the Gesellschaft f\"ur Schwerionenforschung (GSI) \cite{klaft,borneis}. The laboratory value of the transition wavelength was determined to $ 243.87(4) $ nm \cite{klaft}. The uncertainty was dominated by the limitation in the velocity calibration of the fast moving ions. This initial uncertainty of approximately $ 200 $ GHz has been reduced to approximately $ 100 $ GHz in a later analysis \cite{borneis}. 
To overcome this limitation, the deceleration of highly-charged ions and a spectroscopic investigation of trapped ions have been proposed \cite{HITRAP-prop,vogel}. Spectroscopy of highly-charged ions stored in Penning traps is expected to reach accuracies which make tests of bound-state QED calculations significantly more stringent and give access to improved measurement of fundamental quantities \cite{trapassisted-vogel}.\\
At the SPECTRAP experiment at the HITRAP facility \cite{HITRAP-prop} at GSI, systematic measurements on several species of highly-charged hydrogen- and lithium-like ions will improve the spectroscopic resolution by up to three orders of magnitude over previous experiments.
The assembly consists of a cryogenic Penning trap, a frequency stabilized laser for excitation of the hyperfine transition under investigation, and a detection system for emitted fluorescence light \cite{vogel}. The trap will be loaded with highly-charged ions produced at the GSI accelerator facility which are decelerated and delivered by the HITRAP beam-lines and cooled to liquid helium temperatures inside the Penning trap.
For efficient spectroscopy of $^{209}$Bi$^{82+}$ the laser system should produce laser light with a power of several mW at $243.87$ nm and a tuning range significantly larger then the standard deviation of $ 100 $ GHz of the previous measurement. In addition, the laser frequency should be stable to a precision and accuracy comparable to or below the expected Doppler-width of the transition of the ions inside the Penning trap  which is expected to be approximately $ 30 $ MHz.
\section{Laser Design and Stabilisation Scheme}
\label{design}
\begin{figure}
\includegraphics[width=0.7 \textwidth]{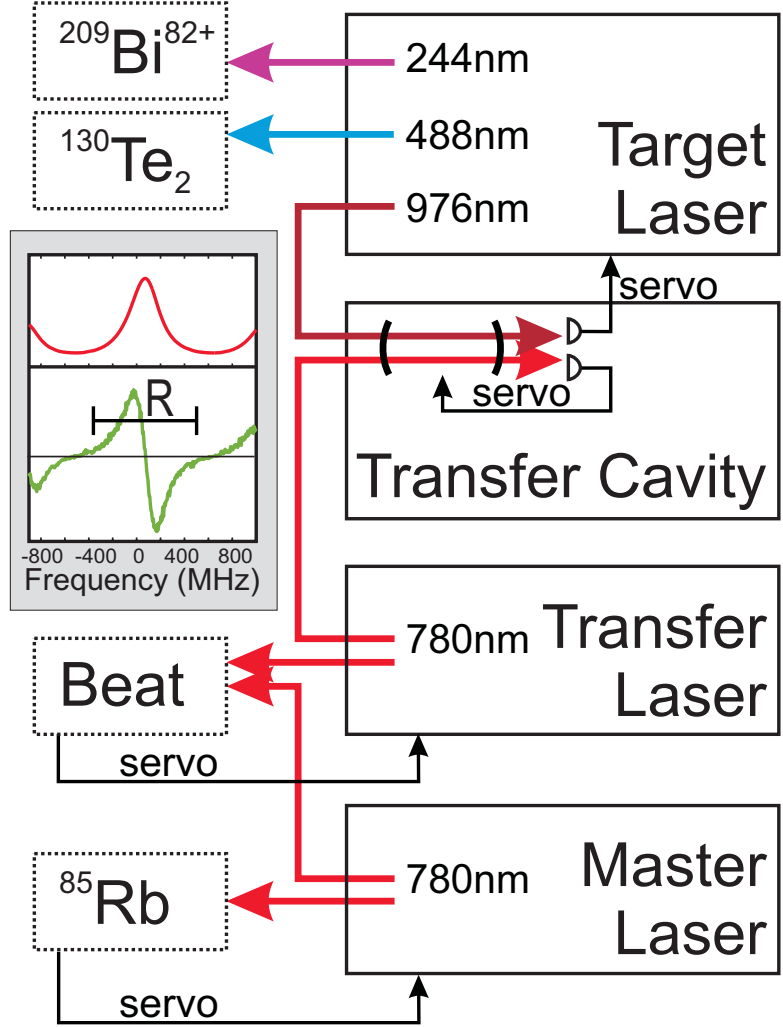}
\caption{Schematic overview of the laser system. The master laser, stabilized to a rubidium transition, serves as absolute frequency reference. The transfer laser is offset-locked to the master laser by an rf beat measurement. The transfer cavity is stabilized to the transfer laser and serves as reference for locking the target laser. Frequency doubling and quadrupling generates light at $ 488 $ and $ 244 $ nm.
In the inset (gray frame) the transmission signal (top) and the calculated dispersive signal (bottom) of the transfer cavity are shown indicating the capture range (R) of about 850 MHz. }
\label{optics-schematics}
\end{figure}

\begin{figure}
\includegraphics[width=0.7 \textwidth]{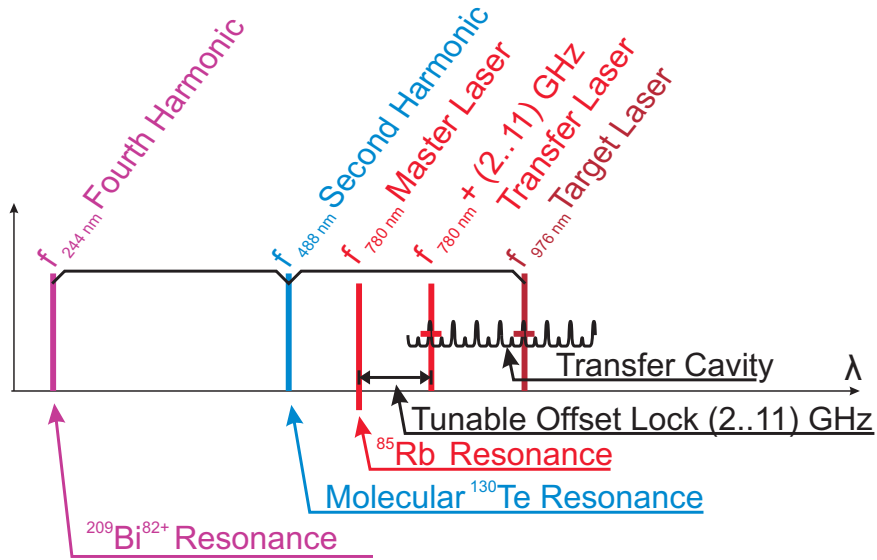}
\caption{Overview in the output frequencies occurring in the frequency stabilized laser system for spectroscopy of hydrogen-like bismuth $^{209}$Bi$^{82+}$.}
\label{frequencies}
\end{figure}

We have developed, built, and characterized a laser systems which fulfils these requirements. The system consists of four major components (Fig. \ref{optics-schematics}): a commercial frequency-quadrupled diode laser system (target laser, Toptica TA-FHG 110) generating light at a wavelength of $ 243.87 $ nm, a pair of offset-locked diode lasers operating at $780$ nm (master and transfer laser), a confocal transfer cavity locking the target laser to the transfer laser at a large wavelength offset, and a set of spectroscopic diagnostics at $ 780 $ nm (rubidium (Rb) absorption cell) and $ 488 $ nm (tellurium ($^{130}$Te$_2$) absorption cell) for characterization of the different components of the frequency stabilisation chain.\\
The target laser produces light around $244$ nm and an output power of up to $ 15 $ mW. This is achieved by fourth harmonic generation of the light of a tapered amplifier with an output power of $ 800 $ mW seeded by an external cavity diode laser operating at a wavelength of $ 976 $ nm. The manufacturer specifies the linewidth of the system to be smaller than $ 4 $ MHz at $ 244 $ nm ($1\mu$s measurement interval). 
The modular setup with two successive frequency doubling stages gives access to three laser fields: light at $976$ nm is used for frequency stabilization, light at $488$ nm is used for frequency diagnostics on $^{130}$Te$_2$ vapour, and light at $244$ nm is used for high-resolution spectroscopy of $^{209}$Bi$^{82+}$.
In order to be able to implement the required large tuning range of several $ 100 $ GHz, the direct stabilization of one of these wavelengths to an atomic or molecular transition is not possible. For that reason, we implemented a stabilization scheme where the fundamental laser output at $ 976 $ nm is locked to a length-stabilized transfer cavity. The transfer cavity serves as a stable reference similar to the one used in \cite{tcav2}, where a cavity is actively stabilized to a polarization stabilized helium-neon laser.
In contrast to this approach, our transfer cavity is stabilized to a tunable diode laser (transfer laser). 
By tuning the frequency of the transfer laser, we can tune the resonance frequency of the transfer cavity and in this manner also the output frequency of the target laser. 
An overview of the involved laser wavelengths is provided by Fig. \ref{frequencies}: The anchor of the stabilization chain is the master laser stabilized to a $^{85}$Rb transition at $ 780$ nm by Doppler-free saturation spectroscopy. The transfer laser is offset-locked to the master laser with a computer-controlled frequency offset in the range of $ 2 $  to $ 10.5 $ GHz with respect to  $780$ nm. 
The transfer cavity is stabilized to the transfer laser, bridges the wavelength gap to $976$ nm, and serves for locking of the fundamental output of the target laser system.
Using a sample-and-hold technique (see Sec. \ref{mopastab}) a scan range of several $ 100$ GHz can be achieved.
Finally, two successive frequency doubling stages generate light at $ 488 $ and $ 244 $ nm.\\
The following sections give a more detailed discussion of the stabilization chain.

\textbf{ }

\subsection{Rb-Stabilization and RF-Offset-Lock}
\label{rbsys}

\begin{figure}
\includegraphics[width=0.7 \textwidth, angle=0]{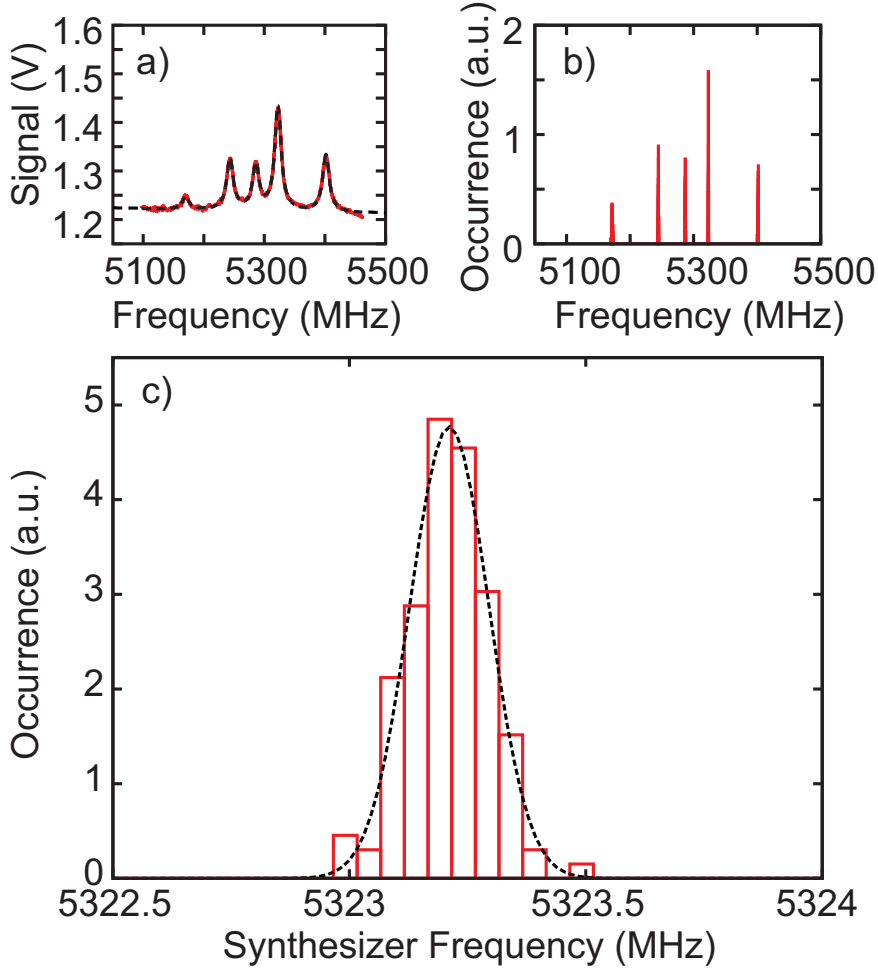}
\caption{(a) Doppler-free absorption spectroscopy on $^{87}$Rb with fitted Lorentzian profiles; (b) Histogram of the extracted centre frequencies; (c) Histogram of the centre frequencies of the $^{87}$Rb $(F=1) \rightarrow  (F'=1, F'=2)$ cross-over transition. The standard deviation ($\sigma_{780 nm} = 83 $ kHz) of the fitted Gaussian function gives a value for the frequency uncertainty.}
\label{hiss_Rb}
\end{figure}

The master and transfer lasers are home-built external-cavity diode lasers operating at a wavelength of $780$ nm. The master laser is stabilized to the $5^2S_{1/2} (F=3) \rightarrow 5^2P_{3/2}(F'=3,F'=4)$ cross-over resonance of $^{85}$Rb by Doppler-free absorption spectroscopy in a mu-metal shielded vapour cell. The stabilization is based on phase-sensitive detection of the saturated absorption signal using a lock-in amplifier with direct modulation of the diode laser's injection current at a frequency of $ 70 $ kHz. The feedback bandwidth of $ 10 $ kHz of the PI servo loop is defined by the time constant of the lock-in amplifier.  
Typical short-term line-widths of the $780$ nm lasers are in the range of $50$ kHz. \\ 
A heterodyne beat-signal between the outputs of master and transfer laser is mixed with the output of an rf-synthesizer with a double balanced mixer. A frequency-to-voltage converter produces an error signal proportional to the frequency offset between the beat and the synthesizer frequency. This error signal is fed to a PI controller for stabilization of the transfer laser frequency.
The transfer laser frequency can be tuned with respect to the fixed master laser frequency with sub-kHz resolution by tuning the synthesizer output.
This allows us to scan the frequency of the transfer laser with high accuracy and in an arbitrary frequency sequence.\\
%
The transfer laser stabilization and tuneability are enhanced using the intra-cavity stabilization scheme described in \cite{scan}. 
Here, the internal cavity of the laser diode consisting of its front and rear surface is mode-matched to an external cavity formed by the rear surface of the laser diode and an external cavity mirror. Mode-matching is controlled by the injection current of the laser diode. With this technique we reach a mode-hop free tuning range of $ 8.5 $ GHz (at $ 780 $ nm), limited by the bandwidth of the beat-note detection.\\
We use an independent Doppler-free absorption spectroscopy on rubidium to characterize the 
performance of the frequency stabilization and to determine the uncertainty of the transfer laser frequency. 
By repeated scans over the Doppler-free resonances of the $5^2S_{1/2} (F=1) \rightarrow 5^2P_{3/2}(F')$ transitions of $^{87}$Rb 
and by fitting Lorentzian profiles to the occurring resonance lines (Fig. \ref{hiss_Rb} (a)) we can extract the centre frequencies of the observed lines. 
The resulting histograms of the centre frequencies of all lines and of the $(F=1) \rightarrow (F'=1, F'=2)$ cross-over transition are presented in Figs. \ref{hiss_Rb} (b) and (c), respectively.
A Gaussian fit (dashed curve in Fig. \ref{hiss_Rb} (c)) gives a statistical uncertainty of $ \sigma_{780 nm} = 83 $ kHz (one standard deviation) for data accumulated over several hours and $  \sigma_{780 nm} = 123 $ kHz for data compiled over several days.

\subsection{Transfer cavity}
\label{CoTCav}
We use a transfer cavity for stabilization of the target laser to the transfer laser with large wavelength difference. The length of the transfer cavity is locked to the transfer laser wavelength using a piezo transducer in one of the cavity mirror mounts.
Our locking scheme, which includes a sample-and-hold circuit for a controlled change of the cavity mode number (see Sec. \ref{mopastab}) is applicable to a wide range of target wavelengths and in addition allows to extend the tuning range of the target laser well beyond the transfer laser tuning range.
The transfer cavity consists of a temperature-stabilized aluminium spacer and low-reflective silver mirrors giving a free spectral range ($ FSR $) of about $ 2 $ GHz.
An aluminium spacer is sufficient for our purpose due to the active length stabilization to the transfer laser. In a later stage switching to Invar, Zerodur, or ULE may be considered.
By changing the cavity length to the next mode number, the $ FSR $ is changed by approximately $ 40$ kHz.
The free spectral range is measured by scanning the laser frequency between absolutely calibrated features of tellurium (Sec. \ref{tespec}) and recording the number of transmission fringes. These measurements resulted in a $ FSR = 2002.345(13)$ MHz.\\
The reflectivity of the cavity mirrors with coatings of $15$ nm Ag and $40$ nm MgF$_2$ was chosen to be $69$ \% at $976$ nm and $70$ \% at $780$ nm, respectively, in order to give a large capture range for the transfer cavity locks. 
The theoretical value for the finesse of $ 8.5 $ is confirmed experimentally by a measured value of $8.24$ ($FWHM=243$ MHz).
Though the large fringe width makes the system sensitive to offset drifts in the electronics, monitoring and correcting these drifts resulted in insignificant contributions to the frequency drifts.
The temperature stabilization of the cavity spacer has small residual thermal variations resulting in piezo transducer stabilization voltage variations on long time-scales. We determined that these variations of approximately $ 30$ V translates to a remaining temperature fluctuation of about $ 0.2 ^\circ$C which corresponds to the expected precision of the temperature stabilization used.\\
The length stabilization of the transfer cavity as well as the stabilization of the target laser to the transfer cavity are done by top-of-fringe locking since it is insensitive to intensity fluctuations.
By modulating the position of one of the cavity mirrors at $ 5 $ kHz we produce the error signals for both stabilization circuits using phase-sensitive demodulation with lock-in amplifiers.
\subsection{Stabilization of Target Laser System}
\label{mopastab}
We stabilize the fundamental wavelength of the target laser system at $976$ nm to the transfer cavity.
A sample-and-hold technique allows us to extend the tuning range of the target laser beyond the tuning range of the transfer laser: 
at the limit of the tuning range, the feedback controller is muted by external control. This causes the control voltage to stay constant ("hold operation") on the last value before muting and the target laser to hold its set-point. The wavelength is then held without feedback, but small wavelength drifts may occur. Next we scan the transfer cavity back by an integer multiple of the $FSR$ and relock the target laser to another cavity fringe.
The large capture range due to the small finesse of the cavity (see inset in Fig. \ref{optics-schematics}) guarantees relocking although the target laser wavelength may have a small deviation from the expected wavelength.
The fringe shifting by sample-and-hold ensures the cavity length on average to stay the same and the $FSR$ to stay on a well defined value.
In addition, the mode-hop free tuning range of the target laser system of approximately $ 40 $ GHz at $ 244 $ nm (limited by the finite tuning range of the doubling and quadrupling cavities) can be fully exploited. Realigning the doubling and quadrupling stages results in a tuning range of the system of more then $ 100 $ GHz in the UV as required.
\section{Results and Discussion}
\label{results}
\subsection{Spectroscopic Measurements on Tellurium}
\label{tespec}
\begin{figure}
\centering
\includegraphics[width=0.7 \textwidth]{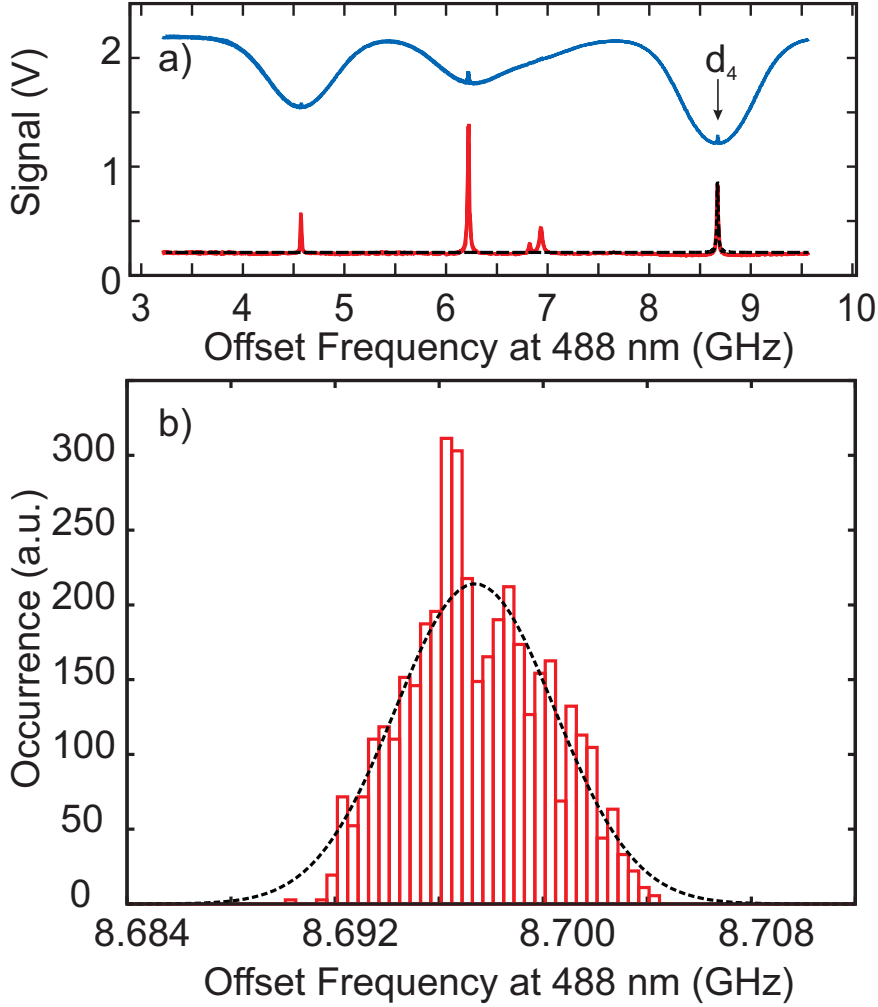}
\caption{(a) Doppler-broadened (blue) and Doppler-free (red) spectrum of tellurium around the d$_4$ reference feature. The dashed line indicates the fitted Lorentzian profile of the d$_4$ line. (b) Histogram of the centre frequencies extracted from repeated measurements of the d$_4$ line. The data were taken over six days resulting in $1300$ individual measurements. The dashed line represents a Gaussian fit with a standard deviation of $\sigma_{488nm}=3.07$ MHz, giving $\sigma_{244nm}=6.14$ MHz.}
\label{hiss-euvar}
\end{figure}

The fundamental frequency of the target laser system is frequency doubled to $488$ nm and then again frequency doubled to $244$ nm by two separate external doubling cavities.
A small fraction of the blue laser light at $488\,\mathrm{nm}$ is diverted for Doppler-free saturation spectroscopy in a vapour cell of tellurium ($^{130}$Te$_2$). 
Similar to iodine,$^{130}$Te$_2$ has many resonance lines that cover most of the visible spectrum,
so that the second harmonic frequency of the target laser can be used to determine the frequency uncertainty of the stabilized laser system.
In addition, high-precision reference lines in $^{130}$Te$_2$ can be used to give absolute accuracy to our laser system.
A $10\,\mathrm{cm}$ $^{130}$Te$_2$ cell similar to the one employed in \cite{Telines d4,Telines d4',Telines a18}, is heated to $500\,^{\circ}\mathrm{C}$
in a tube furnace.
By scanning the laser system, we record the Doppler-broadened and the Doppler-free transition lines of $^{130}$Te$_2$ over a range of several 100 GHz.
We can identify numerous Doppler-broadened lines of the "tellurium atlas" \cite{Telines atlas} and the Doppler-free lines published in \cite{Telines d4,Telines d4',Telines a18,Telines 1088,Telines b2,Telines b1}.
In specific, we scan the known d$_4$ reference line \cite{Telines d4} numerous times (Fig. \ref{hiss-euvar} (a)) and extract the centre frequency of this line, by fitting a Lorentz function to each measurement.
The extracted centre frequencies are plotted in a histogram (Fig. \ref{hiss-euvar} (b)).
A Gaussian fit function describes the distribution well. The standard deviation of the Gaussian is $ \sigma_{488nm} = 3.07 $ MHz (at $ 488 $ nm) for the compiled data from over $ 1300 $ measurements recorded in six consecutive days including a $50$ hour period of continuous laser operation and data taking. 
For the final target wavelength of $ 244 $ nm this corresponds to an uncertainty of 
$ \sigma_{244nm}=6.14 $ MHz which is well below the required uncertainty of $ 30 $ MHz as given by the Doppler broadening of the 
targeted  $^{209}$Bi$^{82+}$ transition.
%
%
 \subsection{Influence of Refractive Index of Air}
\label{pressuretemp}
During the six day measurement run resulting in the data of Fig. \ref{hiss-euvar} (b), we observed a drift in the measured centre frequency of a few MHz, which we attribute to a change in air pressure.  Using this additional information it is possible to further reduce the uncertainty of the target laser output frequency in our open cavity configuration. Ultimately, housing the transfer cavity in a constant-pressure or evacuated environment should eliminate this effect. \\
To gain further insight, we analysed our data in the light of potential temperature and pressure dependences of the target laser wavelength caused by a wavelength dependent variation of the transfer cavity $ FSR$ at $780$ nm and $976$ nm.
A good description of the influence of pressure and temperature on an open transfer cavity is provided by \cite{tcav2}.
The resonance frequency of the transfer cavity for the $ TEM_{00} $ mode is given by 
\begin{eqnarray}
\label{cavglg}
\nu_N = \frac{N c}{2 n d}
\end{eqnarray}
where $ \nu_N $ is the resonance frequency, $N$ is the order of the longitudinal mode, $c$ is the vacuum speed of light, $d$ is the separation of the cavity mirrors and $n$ is the index of refraction of the medium between the mirrors. At the transfer laser wavelength, changes in the refractive index are compensated by keeping the optical path length between the mirrors constant through the feedback loop. 
The stabilized frequency of the target laser, on the other hand is affected by the dispersion of air, especially when large frequency spans are bridged by the transfer cavity.
The refractive index of air under standard conditions $\tilde{n}$ is calculated for $ 780 $ nm and $ 976 $ nm following the procedure given in \cite{nair} relative to the refractive index of vacuum which is 
$\tilde{n}_{vac} \equiv 1$: 
\begin{align}
\tilde{n}(\lambda)-1 & = \nonumber \\ 
\left[ 8342.13  +  \frac{2406030}{130 - {\lambda}^{-2}} + \frac{15997}{38.9 - {\lambda }^{-2}}  \right] & \cdot 10^{-8}
\end{align}
with the wavelength $ \lambda $ (in $\mu $m). The difference in the refractive indices results to $ \tilde{n}_{780}-\tilde{n}_{976}  = \delta \tilde{n} = 9.3065 \times 10^{-7} $.
The dependence of the refractive index of air $ n $ on pressure $p$ (in Pa) and temperature $T$ (in $^\circ$C) is given by
\begin{align}
n(p, T, \tilde{n}(\lambda)) -1 = (\tilde{n}(\lambda) -1) \cdot \frac{1.0413 \times 10^{-5} \cdot p}{1 + 3.671 \times 10^{-3} \cdot T}
\end{align}
according to \cite{nair} 
and can be approximated by a Taylor expansion up to linear order in T around T$_0  =20 ^\circ$C 
through
\begin{align}
\label{taylor}
n(p, T, \tilde{n}(\lambda)) -1& \approx  (\tilde{n}(\lambda) -1) \cdot \nonumber \\ 
( 1.0413 \times 10^{-5} \cdot p ) & \cdot (0.93160 - 3.1860  \times 10^{-3} \cdot \Delta T)
\end{align}
with $ \Delta T = T - T_0 $.
According to equation (\ref{cavglg}), the resulting deviation of the target laser frequency is given by:
\begin{eqnarray}
\Delta \nu  &=&  \nu_N \cdot \frac{\Delta n}{n} \label{errorn}
\end{eqnarray}
For a calculation of the uncertainty in $ n $, the linearised Taylor expansion (\ref{taylor}) is applied to realistic parameters ($ \Delta p \leq 10 \textrm{ hPa} ; \Delta T \leq 0.2 ^\circ \textrm{C}$). Only the difference in the index of refraction $ \tilde{n}_{780}-\tilde{n}_{976}  = \delta \tilde{n} $ which is not compensated by the cavity length stabilization has to be taken into account resulting in:
\begin{align}
\Delta n & (p, T, \delta \tilde{n}) = \sqrt{ (\frac{\partial n} {\partial p} \Delta p)^2  + (\frac{\partial n} {\partial t} \Delta T)^2} ,\\
\textrm{with} \nonumber \\
 \frac{\partial n} {\partial p} \Delta p &=  \delta \tilde{n} \cdot 1.0413 \times 10^{-5} \Delta p \cdot 0.9316  \nonumber \\
  &= 9.0280 \times 10^{-12} \Delta p  \\
 \frac{\partial n} {\partial T} \Delta T &=  \delta \tilde{n} \cdot 1.0413 \times 10^{-5} \cdot  p \cdot (-3.1860 \times 10^{-3} \Delta T) \nonumber \\
  &= -3.0875 \cdot 10^{-9} \Delta T 
\end{align}
using $ p=1000$ hPa.
For the assumed temperature and pressure fluctuations of $\Delta T \leq 0.2 ^\circ \textrm{C}$ and $\Delta p \leq 10 \textrm{ hPa}$, respectively, the contribution of temperature fluctuations can be neglected. Pressure fluctuations result in a change of the stabilized target laser frequency of $ 0.27 $ MHz per hPa at $ 976 $ MHz which scales to  $ 0.54 $  MHz and $ 1.08 $ MHz per hPa at $ 488 $ nm and $ 244 $ nm, respectively. 
For the largest assumed pressure variations of $ 10 $ hPa, the resulting change in the stabilized target frequency of $ 10.8 $ MHz at $ 244 $ nm is still well below the targeted absolute accuracy of $ 30 $ MHz.\\
\begin{figure}
\centering
\includegraphics[width=0.7 \textwidth]{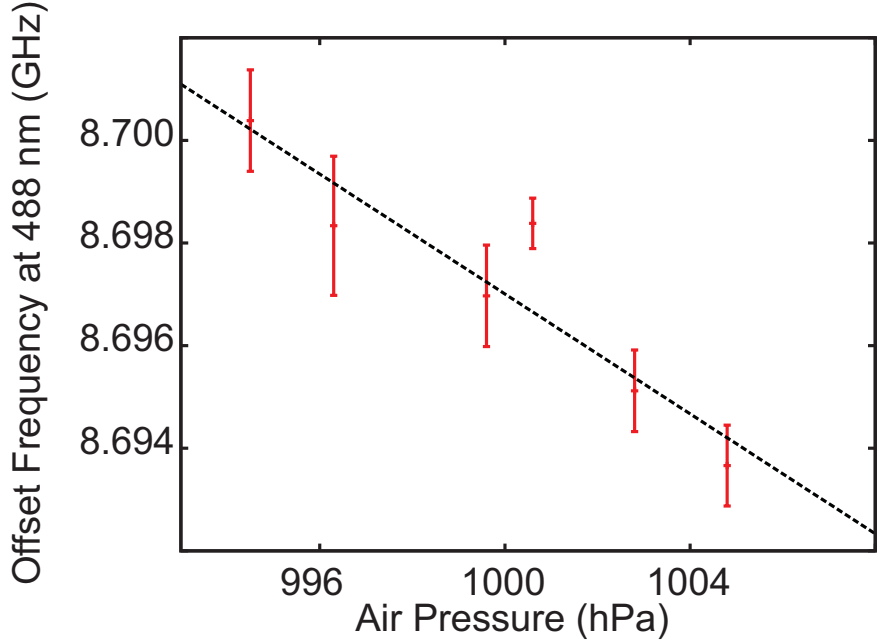}
\caption{
Measured frequency to resonantly excite the tellurium d$_4$-line averaged for each day of data taking as a function of the average air pressure for each day. The linear dependence matches the theoretical prediction based on the pressure dependence of the difference in the refractive indices between $ 780 $ nm and $ 976 $ nm.}
\label{pressureplot}
\end{figure}
We use our data taken over a six day period with significantly varying air pressure to confirm this dependence and to show that knowledge of the air pressure can be used to further reduce the frequency uncertainty of our laser system. Figure \ref{pressureplot} presents the frequency to resonantly excite the d$_4$-line averaged for each day of data taking as a function of the average air pressure reported for the Darmstadt area. We see a clear linear dependence of the d$_4$ frequency on air pressure. Fitting a linear function to the data gives a measured slope of $\frac{\Delta \nu} {\Delta p}$ = $ -0.58(0.11) $ MHz/hPa at $ 488 $ nm. This slope agrees well with the calculated slope of $0.54 $ MHz/hPa when taking to account that an increase in the target laser frequency for increasing air pressure has to be compensated by a decrease in the transfer laser frequency to be still in resonance with the selected tellurium line.
\section{Conclusion}
\label{disc}
In this article, we have presented a laser system for the high-precision measurement of the hyperfine structure of hydrogen-like $^{209}$Bi$^{82+}$.
We have discussed a generally applicable approach for stabilization of a laser system to almost any target wavelength by making use of a rubidium-stabilized laser and a broadband transfer cavity. 
The performance that can be reached depends on the frequency difference between the laser used to stabilize the transfer cavity and the laser to be stabilized to the cavity due to the pressure dependence of the refractive index of air. 
This influence can be further reduced by monitoring the air pressure or by placing the transfer cavity inside a constant-pressure or evacuated environment.
Doppler-free saturation spectroscopy on tellurium can be used to determine the frequency uncertainty of the laser system and to provide reference lines for an absolute frequency calibration of the laser system.

\section*{Acknowlegements}
\label{danke-malte}
We thank M. Schlosser, T. Lauber, J. K\"uber, J. Sch\"utz, S. Tichelmann and the SPECTRAP collaboration for numerous helpful discussions. This work has been supported in part by the BMBF (contract numbers 06DA9019I and 06DA9020I) and the DFG (contract number BI 647/4-1).

\end{document}